\documentclass[conference]{IEEEtran}
\usepackage{cite}
\usepackage{amsmath,amssymb,amsfonts}
\usepackage{algorithmic}
\usepackage{graphicx}
\usepackage{textcomp}
\usepackage{xcolor}
\def\BibTeX{{\rm B\kern-.05em{\sc i\kern-.025em b}\kern-.08em
    T\kern-.1667em\lower.7ex\hbox{E}\kern-.125emX}}
\begin{document}

\title{Distributed VC Firms: The Next Iteration of Venture Capital\\
}

\author{
\IEEEauthorblockN{Mohib Jafri}
\IEEEauthorblockA{\textit{Harvard School of Engineering and Applied Sciences} \\
\textit{M.S. Candidate, Electrical and Computer Engineering}\\
}
\and
\IEEEauthorblockN{Andy Wu}
\IEEEauthorblockA{\textit{Harvard Business School} \\
\textit{Assistant Professor of Business Administration }\\}

}

\maketitle

\begin{abstract}
Using a combination of incentive modeling and empirical meta-analyses, this paper provides a pointed critique at the incentive systems that drive venture capital firms to optimize their practices towards activities that increase General Partner utility yet are disjoint from improving the underlying asset of startup equity. We propose a "distributed venture firm" powered by software automations and governed by a set of functional teams called “Pods” that carry out specific tasks with immediate and long-term payouts given on a deal-by-deal basis. Avenues are provided for further research to validate this model and discover likely paths to implementation. 
\end{abstract}

\section{Introduction}
\subsection{Origin of the Venture Capital Firm}
Though the term has been popularized in recent times since the dot-com boom, the venture capital industry is not a new industry. Many would agree the first venture capital firm was formed in 1946— the American Research and Development Corporation (ARDC) was founded by MIT’s president Karl Compton and Harvard Business School professor Georges F. Doriot \cite{HsuDavidH.2005Ovct}. After a breakthrough success in it’s investment in the Digital Equipment Company, which grew it’s 77\% position for \$77,000 into a \$355 million valuation over 14 years, venture capital solidified its position in the risk capital market. ARDC was a breakthrough proving point in the venture model: that there was an appetite for risk from institutions other than family offices: trusts, university endowments, mutual funds, and insurance companies alike invested in ARDC \cite{GompersPaulA1994TRaF}.

\subsection{Anecdotal Breakthroughs}
Though one can find many anecdotes from the history of venture to prove out the lucrative upside potential of venture, it has proven especially difficult to garner robust, data-driven insights about how venture capital performs overall. This is due to the nature of private markets and the minimal disclosures required to parties not holding a financial interest in the firm. Select groups like Cambridge Associates have been granted privileged access to over 7,300 funds to assess deeper performance metrics beyond regulatory filings, manager surveys, and Freedom of Information Act (FOIA) requests. Their findings were disappointing: for an asset class touted as the most lucrative market to enter, the average VC fund generates a 19\% internal rate of return, while the S\&P 500 has grown at 11\% annually \cite{cambridgeassociates_usvcindex}. Moreover, the spread in venture fund performance tends to exhibit a strong power law: that a small set of firms reap an asymmetric majority of returns for any given vintage year \cite{GompersPaulA1994TRaF}.

\subsection{Financial Mimicry}
While the little data that is available about venture funds call into question how stellar the returns are in actuality, one may also find the origin of the structure of the venture fund as troubling. The traditional “2-20” structure that gives general partners an annual management fee of 2\% of committed capital and a 20\% share of investment profits (carry) date back to the first hedge fund in 1949 by Alfred Winslow Jones, whose stellar 50x return inspired hundreds of new hedge funds and solidified the “2-20” as standard \cite{HewittEd2010LtLH}. Jones picked these numbers on the basis of “Phoenician merchants, who charged one-fifth of profits from a successful journey” \cite{FortadoLindsay2017Hfiq}.

Venture capital’s mirroring of incentives to that of hedge funds was incidental and not founded in a rigorous study of optimal fees. The venture industry therefore fell into the “2-20” system. And with recent research suggesting that venture returns may not be as lucrative as imagined, it is timely to apply a critical lens to the the systemic issues that have brought the industry to where it now lay.

\subsection{Why Venture Capital Matters}
Venture capital removes friction for innovation: it powers the innovation engine by filling the much-needed gap between an entrepreneur at nascent stage in their venture and the point at which they’ve achieved a profitable business— all without entrepreneurs having to risk housing instability, predatory personal loans, etc. While many argue the causality of venture towards innovation— whether one is a leading or lagging indicator of the other \cite{HIRUKAWAMASAYUKI2011VCAI}— one spirit is well-agreed upon: venture capital is in the business of sustaining innovation. This innovation comes from an ambition to tackle unemployment, provide upward mobility to millions, and create impact at astonishingly high scale.

With this motivation in mind, it is of high importance to question the business fundamentals of venture capital. Especially in the presence of data that questions whether the “2-20” standards are as healthy as they’re made out to be, this paper attempts to addresses the following questions: How can we precisely define the problems with the standard venture capital model? Is there a better model that exists to fund innovation via risk capital?

\section{Status Quo:  The Standard Venture Capital Firm Model}

In this section, we outline a simplistic model for the standard technology venture capital fund, capturing the core fee structures that dictate financial incentives for its constituents as well as its fundamental business motions. Then, after an overview of the different methods by which one could measure fund performance, we argue in favor of a set of key performance indicators (KPIs) one ought to prefer. We conclude this section with a detailed critique on the shortcomings of the standard venture model for a majority of its stakeholders.

\subsection{Utility Model}

Let there exist the following parameters:

\begin{itemize}
    \item \textbf{Fund Size} $f \in (0,\infty)$ which specifically refers to paid-in-capital, the net amount that limited partners (LPs) commit to a given fund inclusive of all fees.
    
    \item \textbf{Lifespan} $l \in (0,\infty)$ the lifespan of the fund in years.
    
    \item \textbf{Management Fee} $p \in [0,1]$ the annual management fee charged to the general partners (GPs) for each year of the lifespan of the fund. The denominator is not standard– some take fees on the total of assets under management (AUM), while others take fees solely on the fund size. For modesty, we choose the latter.
    
    \item \textbf{GP Commit} $g \in [0,1]$ the percentage of the fund $f$ that the GPs must personally commit.

    \item \textbf{Carry Fee} $c \in [0,1]$ carry fee percentage, calculated as a percentage of LP distributions that GPs are entitled to after returning the principle.

    \item \textbf{Expected Multiple} $m \in [0,\infty)$ the distributions to paid-in capital (DPI) that a GP expects to return. 

\end{itemize}

Let us assume the following for the standard venture capital firm:
\begin{align*}
    l = 10 \\
    p = 0.02 \\ 
    g = 0.01 \\
    c = 0.20
\end{align*}

This borrows directly from the standard 10 year fund using the “2-20” model with 1\% GP commit. Of course, this model does not account for complexities like time-variant management fees, anchor LP fee minimizing, hurdle rate and clawback provisions, etc. Nonetheless, this model captures the most typical starting point for venture funds \cite{ZiderB1998Hvcw}.

We can therefore express the utility function for the general partners, $U_{gp}$, as:
\begin{align}
U_{gp} = U_{gp_c} + U_{gp_p}
\end{align}

Where is $U_{gp_c}$ denotes the financial carry incentive and $U_{gp_p}$ denotes the performance/management fee incentive to the General Partner. For these two, we have:

\begin{align}
    % \sum_{n=1}^{\infty} 2^{-n} = 1    
    U_{gp_p} = \sum_{L=0}^{l} fp
\end{align}

\begin{align}
    U_{gp_c} = ( (f - U_{gp_p} ) \* m - (f - U_{gp_p} ) ) \* c
\end{align}

The purpose of this model is to make clear the assumptions by which discussions of incentives will be measured. We will revisit this model in a later section.

\subsection{Process Model}
We explicitly state the core functions that any firm will go execute on in order to raise their fund.

\begin{itemize}
    \item \textbf{Capital Commit}: General Partners get commitments to paid-in-capital from Limited Partners via a mandate. This may start at an anchor LP, and move onto high net worth individuals (HNWIs), university endowments, and/or pensions. It is at this time that mandates and thesis are formulated to clarify to LPs how their capital will be invested.

    \item \textbf{Deployment}: General Partners will source, diligence, and close investments on deals that match the constraints of their mandate and thesis. Capital calls will be made to ensure that the fund has enough dry powder on hand to complete the investment.

    \item \textbf{Post Investment Value Add}: Once an investment is executed, GPs have a carry incentive to help their portfolio companies by helping portfolio founders succeed. These value-adds can range from introductions from a GP's network, to hiring, to strategy.

    \item \textbf{Follow-On}: After an agreed-upon percentage of the fund is deployed into initial investments, the mandate may specify capital reserves for follow-on investments into the existing portfolio. This is often where GPs can double-down on better companies.

    \item \textbf{Fund N+1}: Most mandates dictate that performance/management fees are no longer paid after the initial deployment of capital has completed. At this point, fund managers will typically seek to raise their next fund. Whether they raise a similar size fund, graduate to a multistage fund, or shutdown the firm is typically a function of the fund’s latest performance.

\end{itemize}

\subsection{Defining KPIs}

There are many methods to quantitatively benchmark a fund’s performance with varying degrees of tradeoffs. In this paper, we review multiple benchmarks and argue that the TVPI/DPI ratio is the best metric to understanding fund performance.

\begin{itemize}
    \item \textbf{Internal Rate of Return (IRR)}: this metric is a good benchmark for how efficiently a firm put their capital to work. Without loss of generality, this can be understood as a function of how quickly cash on hand increases in perceived in value. However, IRR can be gamed too easily: multi hundred percent gross IRRs are easily attainable for established firms: they can simply call a line of credit to fulfill an investment without having made a capital call to investors. Then, when knowing that their original deal is to be marked up by a later investment, they make the capital call and show incredibly speedy valuation spikes that can rough out slow quarters. More importantly, IRRs are a transient snapshot of a fund's health, which may fluctuate by macroeconomic conditions out of the control of the founder-- none of which accurately predict real cash-on-cash returns. Though IRR is valuable to understand the time-value performance of the fund, it is a shortsighted game to maximize this number every quarter.

    \item \textbf{Total Value to Paid-In Capital (TVPI)}: in lieu of gaming the capital call-timing component as IRR does, TVPI presents an indicator that correlates well to portfolio companies being successful. Whether the venture firm played an active in value creation past capital deployment is out of scope, but this metric indeed controls for how big a bet of LP capital the firm chose to take on a given portfolio company, a valuable proxy for understanding a GP's risk management skills. Yet, looking at TVPI alone can also be gamed-- in fact, it is frequently gamed. A study of 135 private companies that reached unicorn status (greater than  a \$ 1B valuation) found that their post money valuations are 48\% greater than their fair market value \cite{GornallWill2020Svcv}, meaning that TVPIs calculated off of the unicorn valuations can exhibit comparable inflation. Additionally, leading research attributes lesser blame to the macroeconomic conditions at the time of valuation measurement, and more towards informal favors one venture firm may give to another in marking up deals. This phenomena is well-studied at the intersection of venture capital via the gift exchange theory \cite{FerraryMichel2010SoVC}.  The impact is that TVPI points to paper valuations and can be severely discounted when bringing equities to liquidate. 

    \item \textbf{Distributions to Paid-In Capital (DPI)}: this metric is the most fundamentally sound metric: DPI measures how much capital a fund returns (distributions) against how much capital was invested (paid-in capital). Furthermore, gross DPI is net of all fees and carry-- it measures the performance of the fund as it relates to its primary economic buyer, the limited partner/fund investors. This metric is difficult to obfuscate as both distributions and paid-in capital values can be discretely measured and reflects the very beginning and end of a fund's life. However, this metric cannot be obtained until positions are liquidated many years into a fund, which makes it difficult for LPs to understand performance in the short term.

\end{itemize}

We elect to evaluate a venture firm by their TVPI to DPI ratio. This configuration encourages fund managers to maximize a standard metric for value created in the short term, and, knowing that valuation is a lossy measure of value creation, the penultimate burden for the fund is to see how much a fund can maintain and convert their TVPI to DPI.

\subsection{Shortcomings of the Standard Venture Capital Model}

We critique the per annum fee-based, traditional GP-LP venture capital firm model that is standard in technology venture capital, primarily at the seed/pre-seed stage of financing wherein there are little if any performance metrics to evaluate a given investment by. We will show how misaligned incentive structures and vanity metrics have created a non-ideal venture system that incentivizes fund managers disproportionately towards growing their fund and fees as opposed to improving upon the underlying asset they purchase: the value of their portfolio. The impact of this creates an innovation funding engine whose funders' incentives are misaligned from the health of their portfolio businesses.  
\\
\subsubsection{Most venture funds have poor returns}

\begin{figure}[!ht]
\includegraphics[width=\columnwidth]{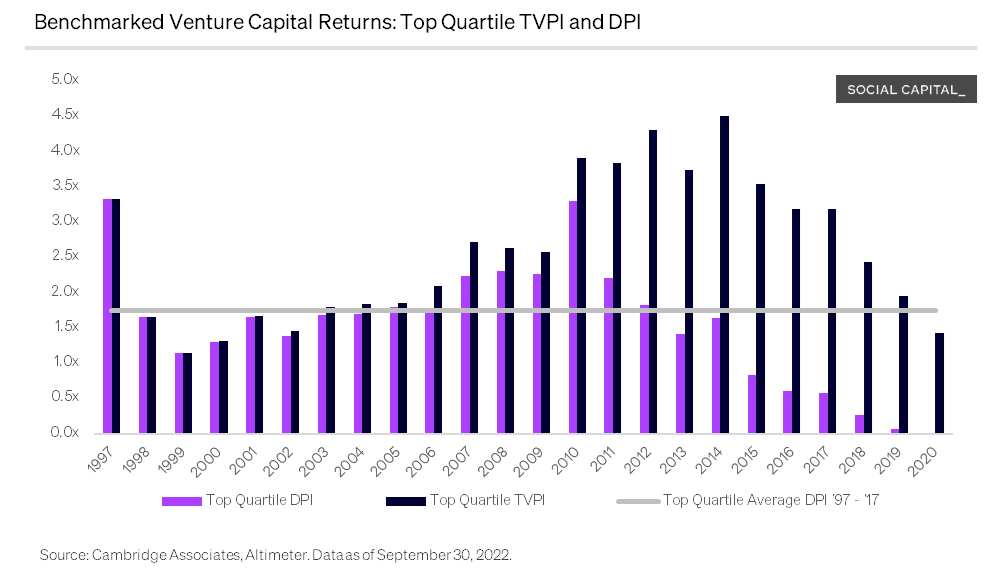}
\caption{Cambridge Associates benchmarks as of 9/30/22. This represents the top quartile venture funds of each relevant vintage. Rates of returns are net of fees, expenses, and carried interest. Cambridge Associates  shows that most funds take at least six years to settle into their quartile ranking. Courtesy of Social Capital. \cite{cambridgeassociates_usvcindex}}
\label{fig:tvpi_to_dpi}
\end{figure}
A 2022 review of vintage year returns since 1997 of the top quartile of venture funds reveal a startling truth (Figure \ref{fig:tvpi_to_dpi}): that even the top performing funds struggle to return a 2x DPI. Especially at the time scale by which TVPIs have reached DPIs (meaning, the fund has fully closed out), a 2x return over roughly 18 years is wildly sub-optimal compared to standards like the S\&P 500.
\\
\subsubsection{Exposing driving incentives}
Recalling Equation 1, we can rewrite the utility function $U_{gp}$ as a  function of  the fundamental inputs, $f$, $p$, $g$, and $l$ as the following: 

\begin{equation}
    \resizebox{.43\textwidth}{!}{$
    U_{gp} = f \cdot (m g - m g p l + p l - g + m c - m c g - m p l c + m p l c g - c + c g)
    $}
\end{equation}

To understand the underlying motives of the GP, we make explicit several assumptions:

\begin{enumerate}
    \item \textbf{GPs wants to maximize their utility.}\par
    Of course, GPs cannot maximize utility to such an extent that they remove any substantial upside from other stakeholders like LPs and founders. In fact, that part can be modeled as another parameter: the probability that LPs or founders will take part in a given setup that maximizes GP utility. For simplicity, we do not consider modeling this dependency as benchmarking LP comfort to risk, reward, timescale, etc. is out of scope.
    
    \item \textbf{GPs do not fully control their DPI multiple.}\par
    This is fair assumption to make as the multiple tracks the effort of both the startup team and the investors, of which over 80\% of venture-backed startups have multiple investors \cite{SorensonOlav2001SNat}. Therefore, we treat a fund's multiple $m$ as a perceived multiple. 
\end{enumerate}

We outline a few strategies for how, for a given expected DPI, a GP can deviate from the standard model described earlier to maximize their utility. This produces the following graph: \\
\\
\begin{figure}[!ht]
\includegraphics[width=\columnwidth]{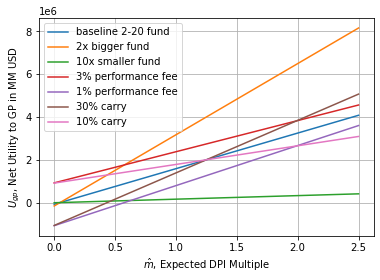}
\caption{The model $U_{gp}$ is given for different perceived DPI multiples that a firm believes they can achieve. Different fund variables are altered against the baseline "2-20" venture firm model.}
\label{fig:driving_incentives}
\end{figure}
\\
Unless a fund performs better than the top quartile that returns 2x, GPs have a clear incentive to increase only two values: management fees or the fund size.
\\
\subsubsection{Fees Over Founders}
As shown by Figure \ref{fig:driving_incentives}, the highest leverage actions that GPs are incentivized to carry out is to increase the size of their fund and raise performance fees, instead of improving the underlying asset-- the equity in founders' startups. Notably, the only metric that strongly correlates to better outcomes for founders is the the carry fee, as that depends on the multiple achieved. However, given that it is historically difficult to produce great returns, most firms will land well below the 2x multiple, and are therefore not incentivized to increase carry. Herein lies the gap between what drives LP-founder utility, and what drives GP utility.

Furthermore, short of breaching fiduciary obligation and violating one's mandate, GPs are protected from all downside: there is a minority of situations as detailed in Figure \ref{fig:driving_incentives} in which GP utility goes below zero for lowest-quartile returns. Even the 1\% GP commit rule is rendered useless as a skin-in-the-game tactic. 

Therefore, in an industry like venture in which the power law dictates returns \cite{GompersPaulA1994TRaF} and the expectation is that most funds will not break into top-decile returns, the highest probability to maximizing a firm’s take-home is raising a larger fund or increasing the management fees. As evidenced by the graph above, any portfolio value-adds combined with market volatility and expectation of a successful exit only further pushes utility to be found at the fee level-- not the founder-level.

A simple refutation to GP's over-optimizing for fees is that firms become fee-taking machines without producing substantial value will cause LPs to stop investing. Unfortunately, LPs will not stop investing due to time skews:  
\\
\subsubsection{Illiquidity Obfuscates Performance}

Recall that the life cycle of the average technology venture fund is 10 years. With a 3-5 year deployment schedule, the fund runs out of dry powder many years before distributions might begin -- that's around the year 10. Looking at Figure \ref{fig:tvpi_to_dpi}, this registers well: funds from 2012 have nearly 3x more in equity that has not converted into returns (or failed). With companies staying private for longer \cite{BegumErdogan2016Gfod}, liquidity is even harder to come by. This long time horizon to liquidity explains why vanity-aligning metrics like IRR (at all time scales) are used in lieu of true KPIs like TVPI to DPI ratio. A fund runs out of dry powder well before they have strong DPIs to show, so they must resort to other metrics to keep the firm alive. 

Therefore, a firm will raise their next fund off of the TVPI of their previous fund. LPs must be convinced that TVPIs will hold-- that they are not inflated paper valuations-- and by the time that Fund I proceeds/distributions start rolling in, when TVPI inflations realize their discount of nearly 50\% to fair market values, firms can easily use rhetoric around the firm's commitment to make Fund II’s TVPI performance exceed Fund I to justify a Fund III.

Therefore, the downstream effect of illiquidity in the private markets is that receiving distributions is so far down the timeline that funds can’t demonstrate true fund performance, and instead raises the next fund off of vanity metrics and rhetoric.
\\
\subsubsection{Constricted Alpha Generation}

As Barnes et. al. describes, institutional LP decisionmaking frameworks tend to look at three evaluation criteria: strong historical data for an indication of returns, a firm's unfair access to the best deals, and a winning strategy for how the capital will be deployed \cite{BarnesSimon2005Iivc}. This proprietary deal flow mechanism is exceptionally difficult to come by when the macro-technological trend is that networks and information are becoming more open, not closed. Look no further than discovery tools like ProductHunt, Pitchbook, Crunchbase, and Twitter. Therefore, to keep a competitive edge, GPs with smaller teams must either upsell their personal networks or hire deal flow associates to convince LPs that they have a stronghold over their mandated market. The traditional fund structure therefore relies on an older worldview that profits off of information asymmetry-- a worldview that is quickly disappearing in the place of movements like build in public \cite{Chapter7Startupstacksunderstandingthenewlandscapeofdigitalentrepreneurshiptechnology}. 
\\
\subsubsection{Poor Risk Management via Rigid Mandates}

Mandates are important. They set the limitations to how GPs can invest, creating an accountability layer for GPs to have a fiduciary obligation to LPs' capital. However, they come with a problem: mandates today are rigid agreements with one-size-fits-all rules. This must be so because of the very nature of a fund being a large sum of alike capital. The overhead to track back particular constraints to particular pieces of capital doesn't scale well. 

This rigidity in mandates create difficult tradeoffs for LPs to navigate: one LP may value a firm's deal flow and investment strategy, but the check size may not be a fit because it doesn't the LP their expected exposure. The core insight is that not all LPs are the same-- they can have massively different risk profile and portfolio desires-- and yet, they sign near-identical mandates to join a firm. The one-size-fits-all mandates force LPs into binary decisions in which tradeoffs have to be made over large sums of capital. 
\\
\subsubsection{Check Sizes are Increasing}
The cost of innovation is decreasing exponentially on three fronts:

\begin{enumerate}
    \item \textbf{Cost of computation:}\par
    cloud computing and Moore's Law have removed the need for startups to spin up their own servers or even design their own chipset: look no further than the TSMCs and the AWS's and their ability to bring the cost of building to record-lows via their infrastructure.
    
    \item \textbf{Cost of development tooling:}\par
    Software is going from a licensure model (e.g. Photoshop) to a freemium model (e.g. Figma). This lowers the barrier to entry to developing real products to requiring even less capital than ever before. 

     \item \textbf{Cost of coordination:}\par
    Project management and sales automation CRM software like Asana and Salesforce respectively enable fewer employees to accomplish projects of greater magnitude and complexity. This decreases the requirement for staffing overhead for any given project. 

\end{enumerate}

And yet, while the cost of launching and scaling a startup is dropping exponentially, why are target check sizes going up? A recent Crunchbase data report finds that the number of startups raising more than \$3M for their first round of capital has doubled from 2015 to 2020 \cite{rowley_2020}. 

Recall one of the two incentive drivers for venture firms to increase their net utility: raising larger funds. With larger funds, one must write larger check sizes for two reasons. First, the administrative overhead in managing a massive set of investments adds strain on both fund administration costs and on a firm's ability to focus on portfolio growth for startups they hold larger positions in. Second, it would be financially irresponsible for a firm to spend any time on a portfolio company where the upside is minimal, where there may be other portfolio companies they have orders of magnitude greater stakes in. Therefore, to minimize these concerns, firms raising larger funds will need to justify larger check sizes with high consistency.

Technology and organizational efficiency are a \textit{leading} indicator that it will be cheaper than ever to start a startup and iterate to product-market-fit. Venture funds are lagging behind on this, and are still set up for capital intensive home runs reminiscent of the early days of computing and the internet. 

In summary, all stakeholders in the traditional venture capital firm suffer from this suboptimal structure. LPs are forced to park their assets into monolithic mandates with high lockup times and rigid risk profiles, only to expect abysmal returns at the end. Founders are minimized to being front-loaded with value-add to inflate TVPI at best (if a fund is early in their cycle), or left fully alone at worst (if a fund is nearly wrapping up). And GPs only get as much deal flow as they have sustained introductions for others' network access-- the "personal brand" of proprietary flow of which is rapidly declining to the closing gap of information asymmetry. 

\section{Venture Capital is Ripe for Change}
\subsection{New Attempts and their Shortcomings}
There have been multiple attempts at innovating upon the venture capital model in it short history of existence. For each of these new models, we will evaluate the value proposition they have as well as their tradeoffs compared to the standard venture capital model.
\\

\subsubsection{Equity Crowdfunding}
Equity crowdfunding platforms like WeFunder and Republic have taken advantage of changing SEC Policies like the JOBS Act that have enabled regulated sales of securities to unaccredited investors. However, there's a clear gap in their success: venture firms will not use these portals because it signals to the LP that the firm doesn't have uniquelt differentiated access to top founders. Furthermore, angels use this in a minority of their investments: only 17\% of all angel investor deal flow identification-- let alone investment closing-- occurs on equity crowdfunding platforms \cite{american_angel}. 
\\
\subsubsection{Investment Syndicates}
Investment syndicates and investment clubs have been popularized by platforms like AngelList Venture. However, the traditional syndication structure presents several key challenges: first, deal flow is constrained, still, to that of the syndicate lead or recommendations they can garner from their personal networks. Second is capital impermanence: the downside of not having a fund is both the inability to guarantee to founders a check and to the syndicate leads a reliable incentive flywheel.
\\
\subsection{Decentralized Autonomous Organizations (DAOs)}
Decentralized autonomous organizations are pooled resource systems that seek to overcome coordination problems through algorithmic governance. Projects like ConstitutionDAO nearly purchased an original copy of the US Constitution, owned by a collective of thousands of invididuals \cite{freeride_daos}. Organizations like MetaCartel Ventures even pioneered legal innovations like code deference to enable LLCs to be managed by code directly from their operating agreement, enabling them to engage in the purchase of real world assets like startup equity  \cite{MCV}. While many DAOs still are exploring the proper governance structure and legal structure that allow their proliferation, the value in DAOs are their ability to act as a substrate for coordinating complex human behavior via algorithmic design. DAOs formalize much of the work of John Learner in uncovering the nature of organizational psychology in venture capital\cite{LernerJoshua1994TSoV}. 
\\
\subsection{The Case for Decentralization in Venture Capital}
The venture capital model can be supercharged by strategic decentralization of certain functions. Note that the term "decentralization" is used to express network-first venture capital and practices that enable collectives rather than oligarchies to preside over the firm, and that the extend and scope of decentralization will vary based on the process in question.

\subsection{Decentralization makes better open markets}
In Markets and Hierarchies \cite{WilliamsonOliverE1983Mah:}, Williamson argues that firms and organizations can be classified into two broad categories: hierarchical and market-based. Hierarchical organizations are characterized by a top-down decision-making structure, where decisions are made by a small group of individuals at the top of the organization and are then passed down to lower-level employees. These organizations often use internal markets, where goods and services are exchanged within the organization, rather than relying on external markets.

Market-based organizations, on the other hand, are characterized by decentralized decision-making and competition among producers. In these organizations, decisions are made by individual producers or consumers, rather than by a central authority. Market-based organizations often rely on external markets to exchange goods and services, and producers must compete with each other to survive.

Williamson argues that the choice between hierarchical and market-based organizational structures depends on a variety of factors, including the complexity of the tasks being performed, the level of uncertainty, and the costs associated with making decisions and coordinating economic activity. He also argues that the optimal organizational structure can change over time, as conditions and circumstances change \cite{WilliamsonOliverE1983Mah:}.

Venture capital can combine open markets with decentralized decisionmaking to increase competition and find better prices across all aspects-- from internal (staffing), to fund structure (fees), to startup valuation. 

\section{Proposed Model: The Distributed Venture Capital Firm}
\subsection{Overview}
\begin{figure}[!ht]
\includegraphics[width=\columnwidth]{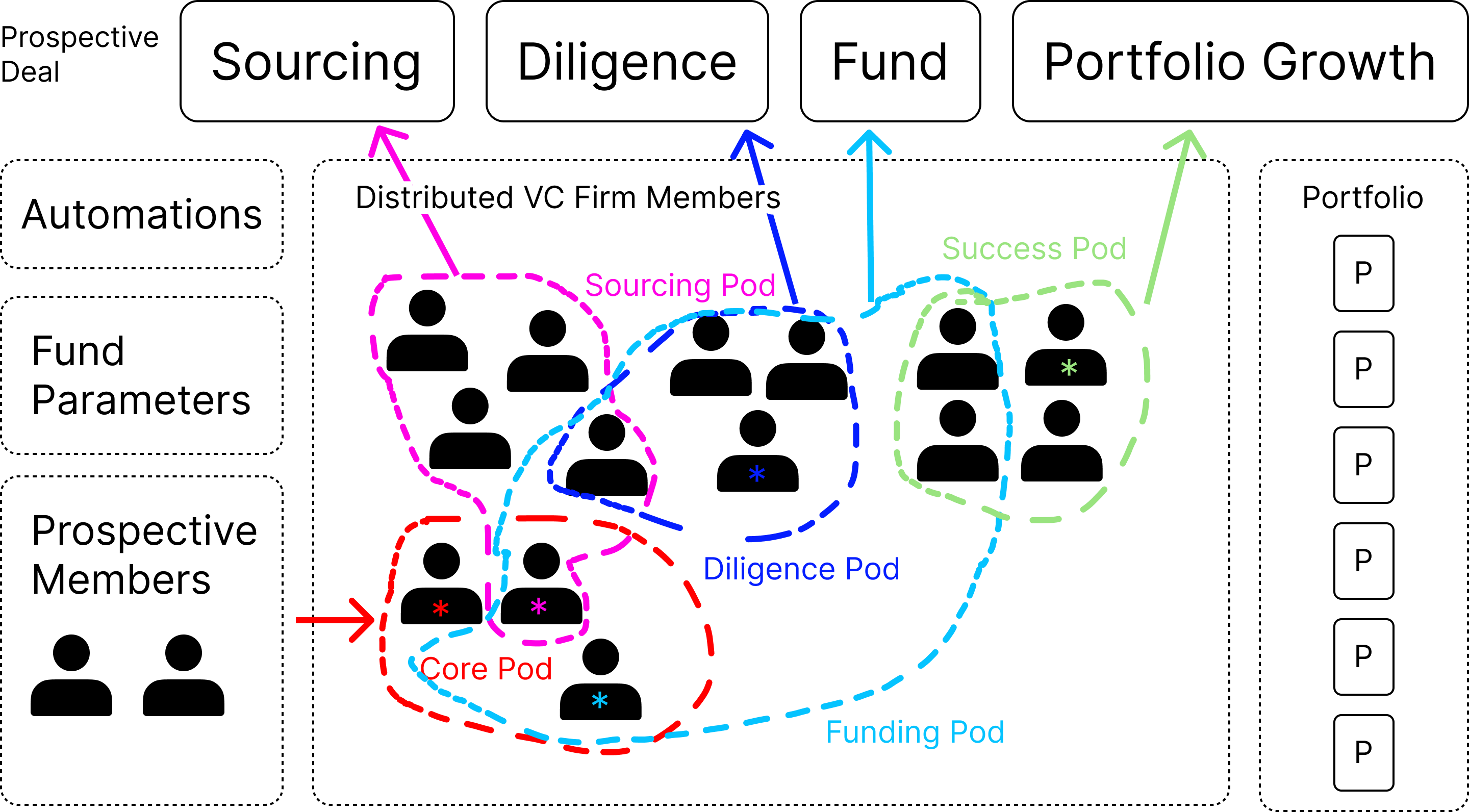}
\caption{A high-level system diagram of a Distributed Venture Capital Firm.}
\label{fig:dvc_diagram}
\end{figure}
We propose a novel "distributed venture capital firm" powered by software automations and governed by a set of functional teams called “Pods” that carry out specific tasks with immediate and long-term payouts given on a deal-by-deal basis. The distributed venture capital firm creates new funding mechanism-- a hybrid structure borrowing from both traditional funds and syndicate structures-- in which anyone in a group can create a deal, layer economics (i.e. carry and fees) on top of the deal at rates pre-determined by the Core Pod, share a common knowledge and experience set for diligencing, and elect to invest manually or via funding automations. In distributed venture capital firms, incentives are no longer centered around the life cycle of a single fund, but rather are fully contained on a deal-by-deal basis with capital permanence maximized by funding automations and coordination overhead minimized by pre-set incentive structures set by the Core Pod. The remainder of this paper will further outline this structure, benchmark its benefits against that of the status quo, and explore some limitations of the distributed venture model. 
\\
\subsection{Formation and Process Flow}
Because the activities of this distributed venture firm are not centered around a unary fund and its emergent incentives, the process flow of this group tends to be quite nonlinear. Instead, each process flow outlined below details a cycle that may be carried out in any order, at any speed, as deemed fit. We therefore describe each functional component of the venture firm to their respective Pods. Note that members may be part of several of these Pods simultaneously.

\subsubsection{Core Pod}
The Core Pod, which looks most like the founding General Partners in traditional firms, tend to the garden and create the flywheels of the firm via incentive alignment. They perform several discrete tasks akin to that of a project manager: 

\begin{enumerate}
    \item \textbf{Community Management}\par
    Core Pod members control the quantity and quality of community members. Core Pod members take the role community moderation, admissions, Pod assignments, etc.
    
    \item \textbf{Dynamic Mandate Setting}\par
    The Core Pod sets default fees (carry, performance) for any deal created, lockup periods, and how fees get shared. For example, if the default fee is a one-time 2\% performance fee and a 20\% carry, the Core Pod may set that the members that sourced the deal from the Sourcing Pod receives 30\% of the performance fee, those who helped understand the deal via the Diligence Pod receives 25\% of the total carry, 30\% of the carry is given to anyone helping portfolio companies via the Success Pod, and the remainder goes to the Core Pod. Note that that these variables greatly affect the extent to which each Pod is incentivized to collaborate, and, given a sufficient number of these distributed venture firms, open markets would determine fair market value and compensation for such activities \cite{KossovskyNir2004UtMT}. 

    \item \textbf{Fund Administration}\par
    There are significant legal, accounting, and tax-related issues that any investing entity must take on. The Core Pod must ensure all backend processes are controlled or delegated to third parties. Accounting software, platform technologies, and the like assist Core Pods in minimizing the staff count required for this, and instead automating much of this with software.     
\end{enumerate}
\subsubsection{Sourcing Pod}
The Sourcing Pod, which looks most like a "deals" team, manages inbound and outbound leads for new deals. Much like a marketing team creates the first point of contact with a company and their customers, so too does the Sourcing Pod become the first point of contact at the distributed venture firm. And, just as most high-touchpoint sales teams receive marketing-qualified leads (MQLs) from the marketing team, so too does the Diligence Pod receive sourcing-qualified leads (SQLs) from the sourcing team \cite{OlivaRalphA2006Ttkl}. These SQLs are typically strong lines of communication with founders who are raising at terms and in an industry conducive to the mandate of the firm, that the firm has not previously sourced before. As such, compensation in the form of finder's fees, carry share, etc. for this group is revolved around the SQL figure, and their subsequent conversion down the dealflow funnel.
\\
\subsubsection{Diligence Pod}
This Pod combines a mix of qualitative and quantitative reduction of the deal/founder into an investment memo for the Funding Pod to inject capital into. From a qualitative perspective, Laura Huang cites a remarkable ability for seasoned angel investors to predict the profitability of a founder's venture to high accuracies in a scientifically understudied phenomena she currently coins as the "gut feel" \cite{HuangLaura2015MtUT}. Members with such ability will prove invaluable given the lack of data a seed stage company can produce to perform diligence over. And while most diligence activities will likely be a function of trust and reputation analysis, traditional quantitative or industry-expert opinions may also provide strong nuance to a deal memo. 
\\
\subsubsection{Funding Pod}
The Funding Pod receives the deal memo and invests in one of two ways: manually (i.e. members investing on a deal-by-deal basis), or via automations (i.e. an LP-like member creates an automation like "I will invest 100k-250k checks into any biotech startup raising at least 5M, max valuation cap 25M, and no more than three of these types of investments per quarter."). To minimize overhead and increase deal velocity, preset funding parameters like default carry and management fee will enable faster coordination.
\\
\subsubsection{Success Pod}
After an investment has successfully been made, a deal becomes a portfolio company and is actively managed by the Success Pod. This Pod is incentivized via some economic mechanism (flat-fee, carry, etc.) to help portfolio companies in any way they can-- from introductions to new customers to advisory services. This team's success may be measured by any combination of the discrete task performed and the valuation increase the collective Pod brought to that company. 
\\

\subsection{Benefits}
\subsubsection{Superior Sourcing Power}
Traditional VCs don’t tend to rely on LPs for network access when it comes to portfolio construction. in fact, the opposite must be pitched: that the GPs possess a vast network for a given investing strategy over than that which the LP has access to, adjusted for investing time overhead. With a wider LP set that includes a spectrum from hyper-inactive LPs (i.e. likely a pension fund that will set some parameters and check in annually for financial statements) to deeply involved HNWIs, GPs can scale their networks past exhausting their first and second degree connections.

This decentralized sourcing makes sense: just as TikTok and YouTube decentralized access to fame which forced Hollywood to end its stronghold over the entertainment industry, so too can distributed venture firms decentralize access to venture capital and end the stronghold of institutional venture capital. With a Sourcing Pod set up with the right incentives for experienced operators and founders to submit QSLs, positive sum games are created in which new people enter the world of venture and the distributed VC firm sees fresh deal flow. 
\\
\subsubsection{Smaller Check Sizes Are Possible}
Distributed venture funds have the ability to write smaller checks for a few reasons. the LP Automations will enable more strategic alignment at more granular level than ever before. Next, fund economics become less important as the predominant math looks more like that of a syndicate investing on deal-by-deal bases. Third, the overhead of managing many smaller investments can be minimized by algorithmic coordination-- from productized fund administration services like SPV platforms to strategic delegation of fund admin tasks and corresponding incentives to those capable of such a task. 
\\
\subsubsection{More Post-Investment Value Add Power than GPs}
The Success Pod can be a disjoint group of people whose primary economic incentive is seeing the portfolio company succeed. While larger firms can't afford to tend to smaller portfolio companies too early, distributed venture capital firms managing less in assets and liabilities have an opportunity to create strong bonds with their portfolio companies and enable access for further allocation down the line.
\\
\subsubsection{Granular Risk Management}
Powered by the distributed venture firm's Automations, LPs can set specific desired holding periods for different assets. For example, if an LP is looking for or is convinced of longer term exposure on healthcare because their balance sheet and conviction aligns with longer return profiles, then they can set longer holding periods. But if they have a shorter position on, say, payments infrastructure in developing countries, and are forecasting the majority of returns being made in the next should half-decade, they would be able to find deals matching those requirements to go to secondary markets earlier. 

This type of risk management across industry exposure is becoming increasingly more important because of the tension between generalist and specialist VC firms. While LPs previously would go to a particular firm to gain exposure into a particular asset class, VC firms increasingly taking the generalist route via becoming a "multi-strategy firm"-- a firm with a team of specialists-- which seems to have no substantial performance difference over unary specialist firms \cite{pitchbook_specialists} \cite{GompersPaul2009SaSE}. The impact is that LPs must now evaluate a much messier slew of theses combined into one firm, versus correlating one firm to one thesis. As such, LP Automations enable LPs to provide a standardized, flexible mandate for all the different types of assets that a distributed venture firm may want to purchase. 
\\
\section{Limitations}
Many of the limitations to the distributed venture capital firm model come in the execution of the structure and processes that dictate. We outline a few of these limitations below and offer refutations where applicable. 
\\
\subsubsection{Limited Follow On Investment}
78\% of early-stage founders found that the three most functional benefits of having investors is upfront capital to run their business, reputation signaling, and having advice and guidance for how to scale \cite{HoppChristian2010Wdvc}. Lower among the hierarchy of needs is an investor's ability to provide follow-on or bridge funding for future rounds. While traditional venture firms are well-suited for follow on checks as mandates may include post-deployment follow on capital, or even investing out of the next fund a firm raises, distributed venture funds have a higher probability not being able to follow founders with more capital into later stages of investment. As a software-powered venture firm, the distributed VC firm would need to have more LP automations that allow LPs to double down on winners-- e.g. "Reserve 40\% of my capital staked into follow-on reserves for any company that reaches a particular set of criteria"). The coordination of such an Automation as well as the discretization of when an LP would want to follow on makes this a nontrivial new feature to the firm. Automations need to achieve the ease of setting larger monolithic mandates with the optional granularity of the distributed VC model. 
\\
\subsubsection{Lack of Regulatory Clarity}
One may argue that started a distributed venture capital firm adds a degree of regulatory uncertainty that makes the risk difficult to justify. There are questions of general solicitation of securities, tax implications, and more that add uncertainty. What's more, the SEC's stance on where liability falls on software-governed funding systems likes The DAO has left many to opt for manual processes over automation \cite{SEC_TheDAO}. While the Automation require further research into where positive precedents may lie, it is clear that the deal execution portion of distributed venture firms mimic the exact structure of syndicating deals via Special Purpose Vehicles (SPVs). SPVs are pass-through legal entities (typically formed as LLCs) in which any number of accredited investors can jointly claim ownership over the entity via investment, after which the entity purchases an asset like a security. This purchase is fully subject to regulatory oversight by the IRS and SEC in the United States-- a Form D exemption is filed with the SEC, and realized capital gains/losses must be reported by in each SPV member's annual tax filings. This legal structure has moved billions of dollars for decades without trouble, easing the concern of at least one function of the distributed venture firm \cite{alv_whatsanspv}. 
\\
\subsubsection{Administrative Fees}

Using a new SPV to invest in every new deal can welcome unreasonably high administrative fees: the marginal admin cost of adding a new portfolio company for a venture capital firm is zero, given that all admin costs are amortized by firm's setup itself. The cost to a distributed venture firm is non-zero, likely on the order of \$10,000, which is the average cost of running an SPV on AngelList \cite{alv_whatsanspv}. To this, we refute on two fronts. First, the annual management fee of 2\% that traditional VC firms charge scales to the size of the fund-- the admin costs do not scale linearly with the size of the fund. Therefore, traditional VC firms are less capital efficient with admin costs because only a minority of management fees will be used for this purpose. Second, there is high likelihood that competition within SPV administration platforms will drive the costs of running an SPV to approach zero. This may come about via Masters-Series structure innovations, the standardization of processes to remove labor overhead, and the race to the bottom for deal volume. Therefore such trends point to the marginal cost of adding a portfolio company to a distributed VC firm to approach zero as well. 
\\
\subsubsection{Membership Quantity}
While more Sourcing Pod members can increase the network reach of the firm, more Diligence Pods can increase the number of industries the firm has competence in, and more Funding Pods can increase the check size and/or number of deals, there are many concerns over scaling to too many members. There can be a diffusion of responsibility, in which there are so many members touching a deal that nobody claims true ownership over it \cite{WallachMichaelA1964Dora}. There can be free rider problems in which members benefit off of the collective's work without making any substantial impact themselves \cite{sep-free-rider}. There is also an attack vector that the Core Pod may not possess the cognitive ability to keep tabs over a large enough organization as popularized by the infamous Dunbar's number \cite{DUNBAR1992469}. 

These are all valid concerns that any firm, distributed or not, must take into account when understanding how to scale their team. What makes distributed venture firms unfairly suited to tackle this is inherent to its design: Pod Leads, codified incentives, and algorithmic coordination. Pod Leads enable the Core Pod to delegate the ownership of sourcing to another individual, should the firm grow large enough to justify it. This is akin to a Head of Marketing. Codified incentives tackle the free-rider problem head-on by giving financial reward not necessarily to the entire Sourcing Pod at large for finding a deal, for example, but rather to share it among the individuals that worked on a given deal. Furthermore, the ability to leverage algorithmic coordination suggests that software communities can scale past Dunbar's number due to the well-defined interfaces that exist. 
\\
\subsubsection{Membership Quality}
Just like any venture firm, the success of the distributed venture firm is highly dependent on the makeup of its membership. Core Pods can visualize how they pick their members on two dimensions: capital availability and execution ability. Therefore, Core Pods must curate a community of capital givers (e.g. institutional LPs, HNWIs), execution focused-individuals (e.g. active founders/operators), and experienced venture capitalists (e.g. fund managers that understand venture dynamics) in order for all functional Pods to succeed. Distributed venture firms have the same quality requirements that traditional VCs have-- it's only the utility/incentive space that changes. 
\\

\section{Conclusion}
\subsection{Contributions}
In this paper, our contributions can be outlined as the following:

\begin{enumerate}
    \item \par
    We distilled the history of venture as it relates to incentive structures, and the historical-- not empirical-- origins of mimicry that have become the status quo today.

    \item \par
    We rigorously determined the best set of key performance indicators to accurately measure the performance of a fund, taking into account that the chosen metric changes with respect to the life cycle of the fund. 

    \item \par We created a mathematical model that captures the incentive dynamics for founding General Partners of a standard "2-20" venture capital fund. 

    \item \par We addressed the multitude of shortcomings in the standard VC model at both a theoretical level (fees drive GP incentives over helping founders, illiquidity, etc.) and the empirical level (even the top quartile of venture funds struggle to return a 2x DPI)

    \item \par We built a case for the strategic decentralization of particular aspects of the venture capital process.

    \item \par We proposed the creation of the "distributed venture capital firm," a hybrid model between venture firms and investment syndicates, powered by software automations and governed by a set of functional teams called “Pods” that carry out specific tasks with immediate and long-term payouts given on a deal-by-deal basis. 
\end{enumerate}

\subsection{Further Research}
This paper moves rapidly from theoretical arguments to empirical arguments, leaving more depth to be explored on both ends. Below, we recommend the following areas where extensions to our research can be applied: 

\begin{enumerate}
    \item \textbf{Deeper fund performance metric benchmarking}\par
    There were many notable fund metrics not included in this paper for the sake of brevity-- net asset value (NAV) before and after distributions, residual value to paid-in capital (RVPI), etc. A more rigorous study of these at the system-modeling level would help understand the data-sparse private markets. 
    
    \item \textbf{Study community makeup}\par
    This paper does not recommend an optimal size for a distributed venture firm. Is it 10, 100, or 1000 members? And what should the makeup of these members look like as the membership scales? 

    \item \textbf{Evaluate implementation feasibility}\par
    Further research that rigorously explores how this this model makes it to market is in order. This can be studied on two fronts: first, for existing VC firms, how do they get to realizing some of the benefits of this new system? We know systemic innovation in even the oldest of VC firms is possible: look no further than Sequoia's bold move to decentralize their own fund \cite{lessin_2021}. Second, for new firms that have yet to exist, what software, legal innovations, etc. need to be created to allow for the proliferation of distributed venture capital firms?

    \item \textbf{Distributed VC Mathematical benchmarking}\par
    Just as we provided clear utility functions for GPs in the standard VC model, so to should there be a model representing incentives for each Pod.
\end{enumerate}

\section{Acknowledgements}
Thank you to Andy Wu for his constant support in refining this paper's arguments and bringing in quantitative research to help understand the intuitions that much of the venture capital industry operates under today. 

Thank you to AngelList (angel.co), Seed Labs (seedlabs.co), MetaCartel Ventures, and the Ethereum community for providing  tools in the venture capital space that enable new emergent dynamics to be invented as we improve the venture capital model.

\bibliography{biblio.bib}{}
\bibliographystyle{plain}

\end{document}